\documentclass[12pt]{iopart}
\usepackage{setstack}
\usepackage{iopams,amsfonts,amssymb}
\usepackage{cite}
\usepackage{graphicx} 
\usepackage{subcaption}
\usepackage{bbm}
\usepackage{bm}
\renewcommand{\P}{\mathcal{P}}
\newcommand{\Z}{\mathcal{Z}}
\def\id{\ensuremath{\mathbbm{1}}}

\newcounter{eqabc}
\def\mynumparts{\refstepcounter{equation}
     \setcounter{eqabc}{\value{equation}}%
     \setcounter{equation}{0}%
     \def\theequation{\ifnumbysec
     \arabic{section}.\arabic{eqabc}{\it\alph{equation}}%
     \else\arabic{eqabc}{\it\alph{equation}}\fi}}

\def\endmynumparts{\def\theequation{\ifnumbysec
     \arabic{section}.\arabic{equation}\else
     \arabic{equation}\fi}%
     \setcounter{equation}{\value{eqabc}}}
\usepackage[hypertexnames=false]{hyperref}

\begin{document}
\title{Hunting for the Quantum Cheshire Cat}
\author{Antonio Di Lorenzo}
\address{Instituto de F\'{\i}sica, Universidade Federal de Uberl\^{a}ndia, Av. Jo\~{a}o Naves de \'{A}vila 2121, 
Uberl\^{a}ndia, Minas Gerais, 38400-902,  Brazil}%
\address{CNR-IMM-UOS Catania (Universit\`a), Consiglio Nazionale delle Ricerche,
Via Santa Sofia 64, 95123 Catania, Italy}
\ead{dilorenzo@infis.ufu.br}
\begin{abstract}
We analyze the proposal of Aharonov, Popescu, Rohrlich, and Skrzypczyk [New. J. Phys.~\textbf{15}, 113015] of disembodying physical properties from particles. 
We argue that a different criterion, based on the cross-average 
$\langle \mbox{`}cat\ somewhere \mbox{'}\times \mbox{`}grin\ somewhere\ else\mbox{'}\rangle$ should be used to detect the disembodiment, 
rather than the local averages $\langle \mbox{`}cat\ somewhere \mbox{'}\rangle$ and $\langle\mbox{`}grin\ somewhere\ else\mbox{'}\rangle$. 
Here, the exact probability distribution and its characteristic function are derived for arbitrary coupling strength, preparation and post-selection. This allows to successfully hunt down the quantum Cheshire cat, showing that it is a consequence of interference, that it is present also for intermediate-strength measurements, and that it is a rather common occurrence in post-selected measurements. 
\end{abstract}
\section{Introduction}
Weak measurement followed by post-selection \cite{Aharonov1988} is a powerful inference technique, allowing e.g. to reconstruct the unknown wavefunction of a system \cite{Lundeen2011}, or the density matrix \cite{Hofmann2010,DiLorenzo2011a,DiLorenzo2013a,Lundeen2012,Wu2013,DiLorenzo2013f}. 
Perhaps ``measurement'' is a misleading term, since due to the weak interaction between system and meter 
one cannot infer substantial information from a single trial. However, the statistical analysis of the post-selected data --- 
which generally is limited to the average readout but could be extended to the full statistics \cite{DiLorenzo2008,DiLorenzo2012a,DiLorenzo2012e} ---
allows to extract information about the system that is not trivially recovered from standard strong projective measurements. 
The coherent quantum nature of the meter was shown to be of the essence for the peculiar 
amplification of the weak measurement \cite{Duck1989}. Several experimental works 
have focused on signal amplification \cite{Ritchie1991,Hosten2008,Dixon2009,Gorodetski2012}. 
Currently, there is some debate about the usefulness of weak measurement for amplification, 
\cite{Brunner2010,Feizpour2011,Kedem2012,Struebi2013,Knee2013a,Tanaka2013,Knee2013b,Ferrie2013}. 

Recently, Ref.~\cite{Aharonov2013} proposed a way to realize a ``Cheshire cat'' by using joint weak measurements of commuting observables of a photon. The claim is that the grin of the cat (the polarization of the photon) is at one arm of the interferometer, while the cat (the photon itself) is at the other arm. 
Here, we argument that the criterion used for detecting the presence of such a phenomenon, namely 
that the average outputs of the presence and polarization measurements are both 1, is inadequate, 
and we propose using a signed cross-average of the outputs as an indicator of the Cheshire cat, with the 
sign being provided by the post-selection succeeding or not. 

In addition to this, we show that such a disembodiment occurs almost always provided that 
(1) the system is both prepared and post-selected in states that are (partially) coherent superpositions 
of distinct trajectories (in particular, when no post-selection is made, the post-selected state is the identity, i.e. the totally incoherent state and no Cheshire cat shows), and (2) the measurement is not too strong (it does not need to be weak, however).
To this goal, we study the exact statistics of the measurement readouts. 
The main and more general results of the present paper 
are resumed by Eqs. \eref{eq:gengenpost}
--\eref{eq:gengenavs}. The relevant quantities are defined in Eqs.~\eref{eq:rho0}, \eref{eq:aux}, \eref{eq:charfunc0},\eref{eq:gengenwval}, \eref{eq:gengenaddwval}. 
%
\section{Review of the proposed setup} 
\begin{figure}
\centering
\includegraphics[width=4in]{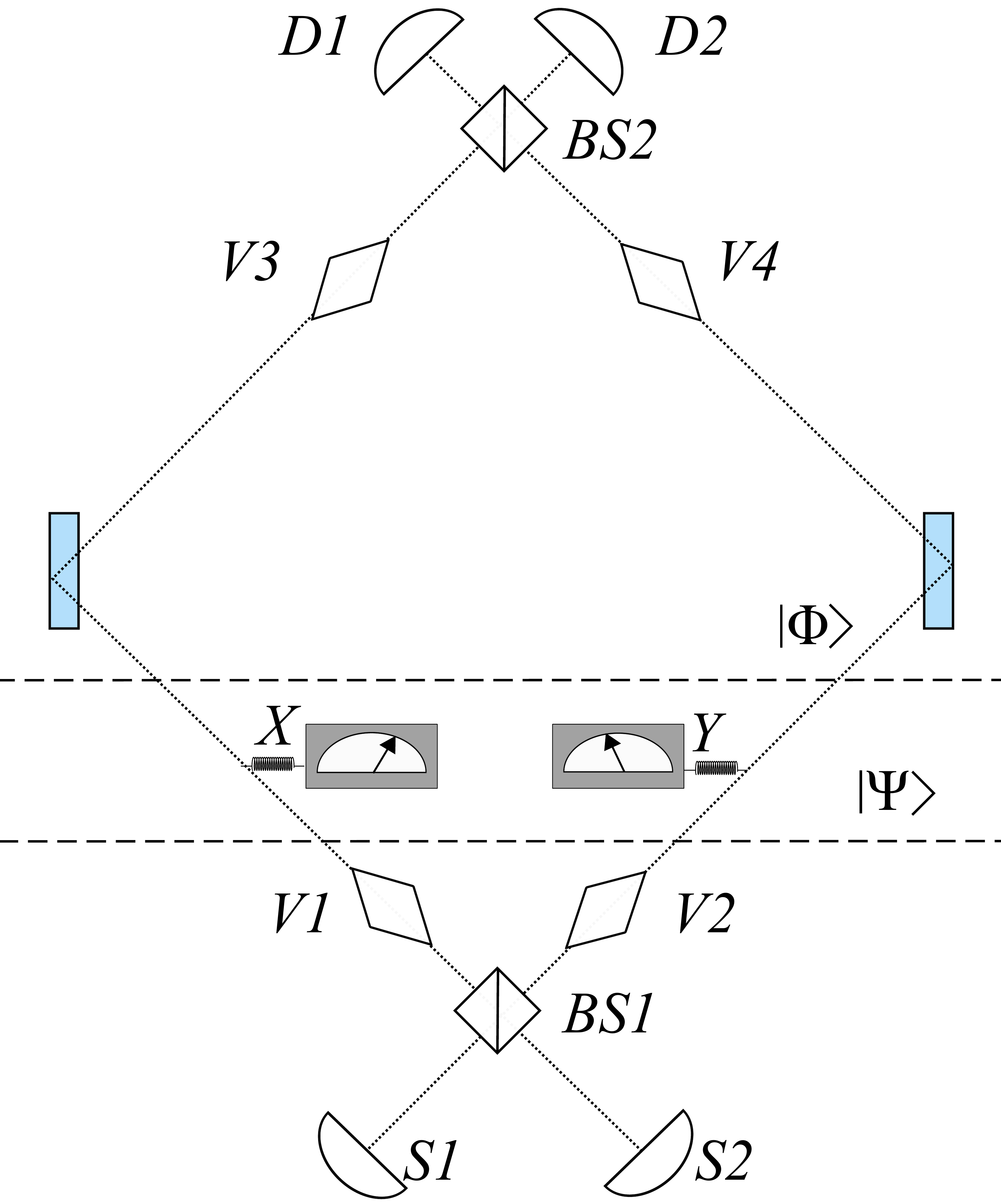}
\caption{\label{fig:setup} The setup. The source $S1$ emits a photon with linear polarization along an arbitrary axis, chosen as the horizontal axis. An asymmetric beam-splitter $BS1$ splits the photon into a left-travelling and a right-travelling component. Local unitary transformations $V1$ and $V2$ are applied to each component. The state of the photon, just before interacting with the meters, is $|\Psi\rangle$, and it can be made arbitrary by appropriately choosing the 
transmission and reflection amplitudes of $BS1$ and the transformations $V1$ and $V2$. 
After the interaction, each path is reflected by a mirror and it undergoes a further transformation $V_3$ or 
$V_4$. Then, a second asymmetric beam-splitter $BS2$ either transmits the beam to the right, where a detector $D2$, able to discriminate between two orthogonal polarizations, is present, or to the left, where there is a detector $D1$, which is also able to discriminate between two orthogonal polarizations of the photon, giving a different signal for a different polarization. Finally, the pure state $|\Psi\rangle$ can be replaced by an arbitrary mixed state $\rho_\mathrm{i}$ 
by judiciously alternating the source $S1$ with the source $S2$ and by having an appropriate 
fraction of the emitted photons  
with vertical rather than horizontal polarization. Analogously, by selecting an appropriate mixture 
of the post-selection measurements when $D1$ reveals a vertically polarized photon, or when $D2$ 
clicks revealing either a horizontally or vertically-polarized photon, one can have a post-selection in a mixed state $E_\mathrm{f}$.}
\end{figure}

With reference to Figure~\ref{fig:setup}, from bottom to top, a horizontally polarized 
photon impinges on a non-polarizing beam splitter (BS1), with asymmetric reflection and transmission, 
so that the new state of the photon is 
\begin{equation}
|\Psi\rangle = r_1|L,H\rangle+t_1|R,H\rangle ,
\label{eq:in0}
\end{equation}
with $L,R$ referring to the spatial part of the photon propagating, respectively, in the left and right arm, 
and $r_1,t_1$ the reflection and the transmission amplitude of the beam-splitter.  
In each arm of the interferometer, an appropriate combination of polarization rotators and 
quarter-wave plates transforms the local state into an arbitrary polarization state, 
\begin{equation}
V_1 r_1|L,H\rangle+V_2 t_1|R,H\rangle =|\Psi\rangle, 
\end{equation}
with $V_1,V_2$ unitary matrices in the polarization subspace. 
Then, in the left arm a meter measures the presence of the photon (the ``cat'')
characterized by the observable 
\begin{equation}
\Pi_\mathrm{L} = 1 |L\rangle \langle L| + 0 |R\rangle \langle R|  , 
\end{equation}
while in the right arm a meter measures its polarization (the ``grin'') along an arbitrary direction $\mathbf{n}$ of the Bloch sphere (e.g., it can establish left-handed or right-handed polarization, etc.),  
\begin{equation}
\sigma_\mathrm{R} =   |R,+\rangle \langle R,+| -  
|R,-\rangle \langle R,-|  , 
\end{equation}
with $|+\rangle$ and $|-\rangle$ the eigenstates of polarization in the direction $\mathbf{n}$. 
In the von Neumann measurement scheme,  the time-evolution operator during the measurement is 
\begin{eqnarray}
\fl \ U= \exp{\left[\rmi \left(a\hat{P}_{X} \Pi_L+b\hat{P}_{Y} \sigma_R\right)\right]} 
\nonumber
\\
= 
\left\{1+\left[\exp(\rmi a \hat{P}_\mathrm{X})-1\right]\Pi_L\right\}
\left\{\Pi_L+\cos(b \hat{P}_\mathrm{Y}) \Pi_R+\rmi \sin(b \hat{P}_\mathrm{Y})\sigma_R\right\}
\nonumber
\\
= 
\exp(\rmi a \hat{P}_\mathrm{X})\Pi_L+\cos(b \hat{P}_\mathrm{Y}) \Pi_R+\rmi \sin(b \hat{P}_\mathrm{Y})\sigma_R
,
\label{eq:timevol}
\end{eqnarray}
with $\Pi_R=|R\rangle\langle R|$ the projector on the right arm of the interferometer, and 
the readout variables being the conjugate variables of $\hat{P}_\mathrm{X}$ and $\hat{P}_\mathrm{Y}$, i.e. the positions $\hat{X}$ and $\hat{Y}$. 
The meters are assumed to be initially uncorrelated between them (otherwise they could give false positives), and with the system (otherwise it would be problematic to say we have a measurement at all), 
so that the density matrix before the interaction is 
\begin{equation}
\hat{\rho} = |\Psi\rangle\langle \Psi| \otimes \hat{\rho}_\mathrm{X} \otimes \hat{\rho}_\mathrm{Y} .
\label{eq:rhoin}
\end{equation}
After the interaction with the meters, the two paths pass through a different combination of 
polarization rotators $V_3$ and $V_4$, then they enter a second 
beam splitter ($BS2$), that is engineered to take the state $r_2|L\rangle+t_2|R\rangle$ to the left 
$|L'\rangle$, and 
the orthogonal state $t_2^*|L\rangle-r_2^*|R\rangle$ to the right $|R'\rangle$ (in which case a photon counter $D2$ will click). 
Finally, a polarization-sensitive detector $D1$ reveals whether there is a horizontally-polarized photon or a 
vertically-polarized one.  
This way, if the photon detector $D1$ reveals a horizontally polarized photon, we can be sure that the state of the photon immediately after the measurement  is 
\begin{equation}
 |\Phi\rangle = V_3^\dagger r_2^*|L,H\rangle+V_4^\dagger t_2^*|R,H\rangle . 
\label{eq:postate}
\end{equation}
A simple way for understanding that this is the post-selected state is to consider an event in which the 
detector $D1$ measures a horizontally-polarized photon, and then visualize the 
backward-in-time process: the detector $D1$ now acts as a source, and it emits a photon with horizontal polarization; this photon impinges on the beam splitter $BS2$, and it splits into the state $r_2^*|L^*,H\rangle+t_2^*|R^*,H\rangle$, 
where $|L^*\rangle$ is the time-reversal state of $|L\rangle$, etc; then, the photon undergoes two local unitary transformations, $V_3^\dagger$, $V_4^\dagger$, which gives the state \eref{eq:postate}. 
The photon then interacts with the two meters, it impinges on the beam-splitter $BS1$, and is finally detected by $S1$, which acts as a photon detector sensible only to horizontally-polarized photons. 
We remark that, in the time-reversed process, only the state of the photon is reversed, while the meters 
are assumed to be in the same initial state $\rho_\mathrm{X}\otimes\rho_\mathrm{Y}$, uncorrelated to the state of the photon. 
This consideration is the key to resolve the arrow of time problem: as observers who have direct access only to  
our states of consciousness, we postulate at every instant of time that the joint state of ourselves and the 
rest of the universe takes the form $\rho_\mathrm{subj}\otimes\rho_\mathrm{obj}$. This subjective 
reduction introduces the time asymmetry that cannot be explained by merely considering the laws of physics.  

Finally, by analyzing data corresponding to the detector $D1$ revealing a horizontally polarized photon, 
it is possible to access the conditional probability of observing readouts $X,Y$ in the meters, given that 
the system was prepared in $|\Psi\rangle$ and post-selected in $|\Phi\rangle$. 
We note that, by considering an appropriate mixture of input states, the initial state of the system can 
be made to coincide with an arbitrary density matrix $\rho_\mathrm{i}$, while a similar procedure in 
the post-selection leads to the mixed post-selected state $E_\mathrm{f}$.  
%

\begin{figure}
\centering
\begin{subfigure}[b]{0.4\textwidth}
\includegraphics[width=\textwidth]{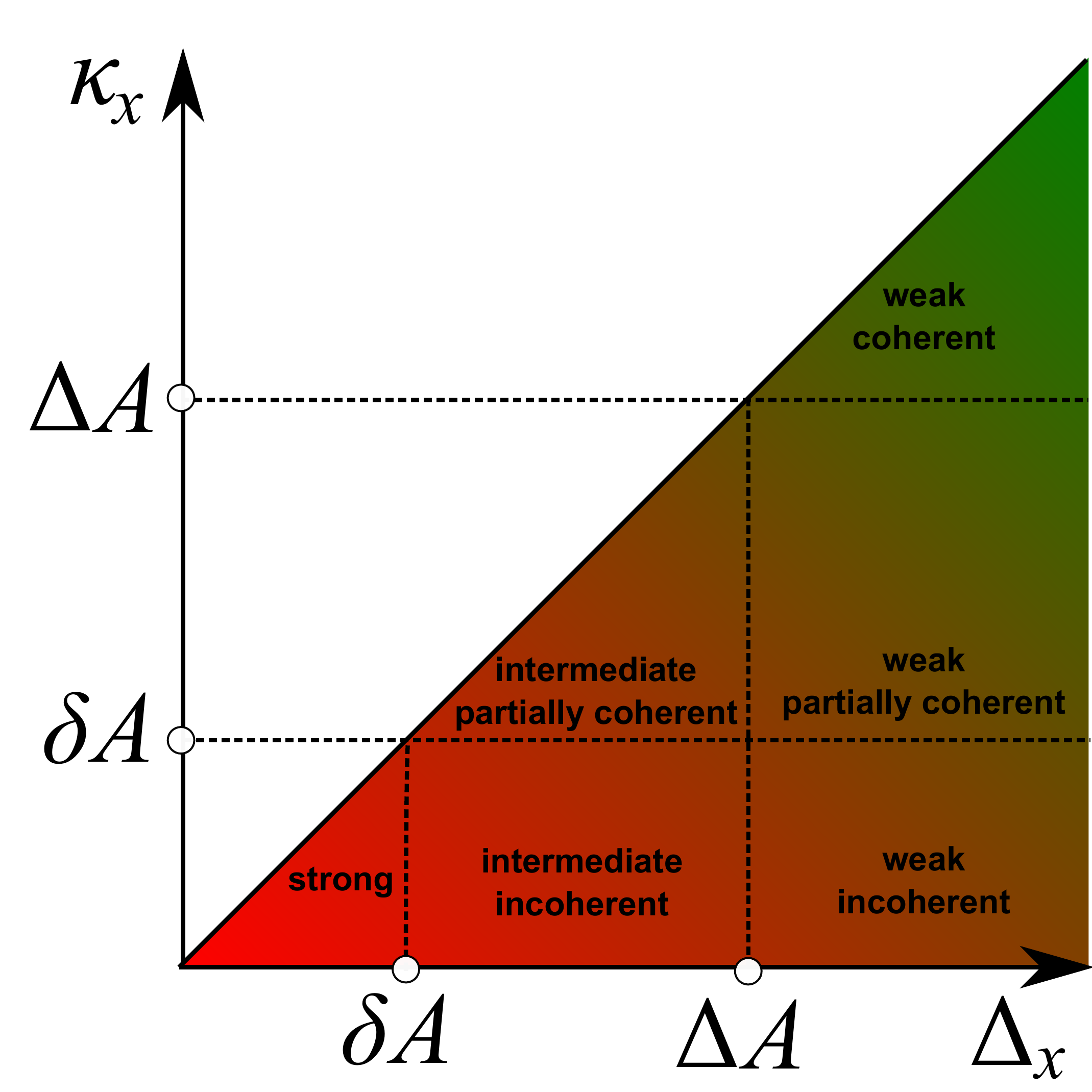}
\caption{}
\label{fig:regimes1}
\end{subfigure}
\qquad
\begin{subfigure}[b]{0.4\textwidth}
\includegraphics[width=\textwidth]{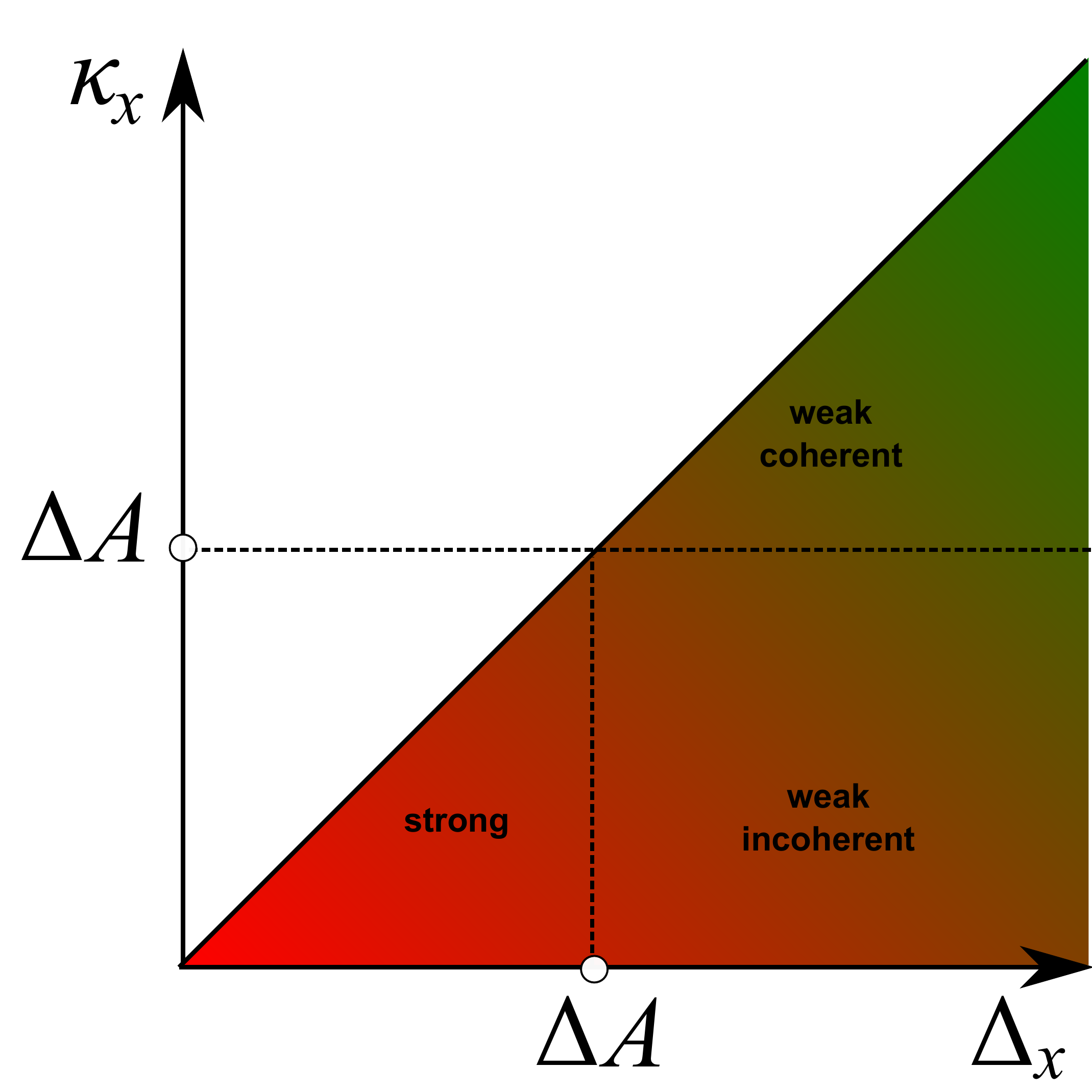}
\caption{}
\label{fig:regimes2} 
\end{subfigure}
\caption{(\subref{fig:regimes1}) The regimes for the measurement of a general 
observable $\hat{A}$. $\Delta_x$ represents the initial spread in the readout variable of the meter, measured in units of the coupling constant, and $\kappa_x$ represents the coherence scale of the meter. 
(\subref{fig:regimes2}) The measurement regimes for a two-valued observable $\hat{A}$, so that $\delta A=\Delta A$, the case considered throughout this paper.}
\label{fig:regimes}
\end{figure}

In order to understand better the various measurement regimes, we consider 
a measurement of a single observable $\hat{A}$ having an arbitrary spectrum $\{A\}$, with a von Neumann 
interaction leading to the time-evolution operator $U=\exp(\rmi \hat{A}\hat{p})$. 
We define the minimum spacing of the eigenvalues of $\hat{A}$, 
$\delta A=\mathrm{min}_{A\neq A'}\{|A-A'|\}$, and their range 
 $\Delta A=\mathrm{max}\{|A-A'|\}$. We also define as $\Delta_x$ the initial uncertainty of the readout 
variable of the meter, $\hat{x}$, 
and with $\kappa_x=(2\Delta_{p})^{-1}$ the coherence length, where $\Delta_{p}$ is the initial uncertainty 
of the variable $\hat{p}$. 
In Figure~\ref{fig:regimes1} the various regimes of measurement are illustrated. 
The intermediate regime was recently shown to be of great interest, for joint measurement \cite{DiLorenzo2011}, for tomography \cite{DiLorenzo2013a,DiLorenzo2013f}, and also for the violation of 
the Heisenberg noise-disturbance relation \cite{DiLorenzo2013c}.  
For an observable having 
only two eigenvalues, however, $\delta A=\Delta A$, so that the intermediate regime is absent, as shown in 
Figure~\ref{fig:regimes2}. 
 
\section{Pure preparation and post-selection}
We shall start with the simpler case of a preparation and a post-selection in a pure state. 
The initial state of the system $|\Psi\rangle$
and its post-selected state $|\Phi\rangle$ are otherwise arbitrary. 
The statistics can be expressed in terms of the weak values
\begin{equation}
L_\mathrm{w} = \frac{\langle \Phi|\Pi_\mathrm{L}|\Psi\rangle}{\langle \Phi|\Psi\rangle},
\qquad 
R_\mathrm{w} = \frac{\langle \Phi|\Pi_\mathrm{R}|\Psi\rangle}{\langle \Phi|\Psi\rangle},
\qquad
\Sigma_\mathrm{w} = \frac{\langle \Phi|\sigma_\mathrm{R}|\Psi\rangle}{\langle \Phi|\Psi\rangle}.
\end{equation}
As $\Pi_L+\Pi_R=1$, $L_\mathrm{w}+R_\mathrm{w}=1$, so that there are 4 independent real parameters.  

From here on, we shall work with normalized variables for the meters $x=X/a$ and $y=Y/b$. 
For conciseness, in the following we shall omit the common prior conditions in 
the probability, so that $\P\{\Phi,x,y|\Psi,\rho_\mathrm{X},\rho_\mathrm{Y}\}=\P\{\Phi,x,y\}$, etc. 
The joint probability of observing the outputs $x,y$ and making a successful post-selection is 
found by applying Born\rq{}s rule to the density matrix of system and meters evolved from \eref{eq:rhoin}  
through \eref{eq:timevol}, 
\begin{equation}\label{eq:Born}
\P\{\Phi,x,y\} = \langle \Phi,x,y| U \rho U^\dagger |\Phi,x,y\rangle .
\end{equation}
Let $\P_\mathrm{X}(x)\equiv \langle x |\hat{\rho}_\mathrm{X}|x\rangle$ and $\P_\mathrm{Y}(y)\equiv \langle y |\hat{\rho}_\mathrm{Y}|y\rangle$ the initial distribution of the readouts. 
In the following, we use the shorthand notation $x_-=x- 1$ and $y_\pm=y\pm1$. 
We have then 
\begin{eqnarray}
\fl \P\{\Phi,x,y\} = 
\left|\langle \Phi|\Psi\rangle\right|^2 
\biggl\{
|L_\mathrm{w}|^2 \P_\mathrm{X}(x_-) \P_\mathrm{Y}(y)
+\frac{1}{4} \sum_\pm |R_\mathrm{w}\pm\Sigma_\mathrm{w}|^2  \P_\mathrm{X}(x)  \P_\mathrm{Y}(y_\mp)
\nonumber 
\\
\nonumber
+\frac{1}{2}\sum_\pm \left[L_\mathrm{w}^* (R_\mathrm{w}\pm\Sigma_\mathrm{w})  \rho_\mathrm{X}(x_-,x)\rho_\mathrm{Y}(y,y_\mp) + c.c.\right]
\nonumber
\\
+\frac{1}{4}\sum_\pm (R_\mathrm{w}\pm\Sigma_\mathrm{w}) (R_\mathrm{w}^*\mp\Sigma_\mathrm{w}^*) \P_\mathrm{X}(x) \rho_\mathrm{Y}(y_\mp,y_\pm)
\biggr\}.
\label{eq:jointprobg}
\end{eqnarray}
It is essential to note how the third term in the right hand side of \eref{eq:jointprobg} contains 
a term shifting simultaneously the arguments of both $\rho_\mathrm{X}$ and $\rho_\mathrm{Y}$. 
This is a signature of interference between the two arms of the interferometer. 
The fourth term, on the other hand, describes interference between the two outputs of the polarization 
detector, and thus it is of less importance for the present study. 	

The initial state of the meters is assumed for simplicity and for definiteness a Gaussian with zero average. 
We do not assume pure states, but 
\mynumparts
\label{eq:rho0}
\begin{eqnarray}
\rho_\mathrm{X}(x,x') = \frac{\varepsilon_\mathrm{X}}{\sqrt{2\pi}}\exp{\left[-\varepsilon_\mathrm{X}^2(x+x')^2/8 -\tilde{\varepsilon}_\mathrm{X}^2 (x-x')^2/8\right]},\\
\rho_\mathrm{Y}(y,y') = \frac{\varepsilon_\mathrm{Y}}{\sqrt{2\pi}}\exp{\left[-\varepsilon_\mathrm{Y}^2(y+y')^2/8 -\tilde{\varepsilon}_\mathrm{Y}^2(y-y')^2/8\right]}.
\end{eqnarray}
\endmynumparts
We remark that $1/\varepsilon_\mathrm{X}=\Delta_x=\Delta_X/a$ is the initial uncertainty of the variable $x$, while $\tilde{\varepsilon}_\mathrm{X}=2\Delta_{p_x}$ is twice the initial uncertainty on the conjugate variable $p_x$. 
Furthermore, $\kappa_x=1/\tilde{\varepsilon}_\mathrm{X}$ represents the coherence scale of the pointer variable $x$, 
as it establishes the scale for the vanishing of the off-diagonal elements of $\rho_\mathrm{X}$ in the $x$-representation. 
Because of the uncertainty principle, $\tilde{\varepsilon}_\mathrm{X}\ge \varepsilon_\mathrm{X}$, and, equivalently, 
$\Delta_x\ge \kappa_x$. Analogously, 
$\tilde{\varepsilon}_\mathrm{Y}\ge \varepsilon_\mathrm{Y}$ and $\Delta_y\ge \kappa_y$. 
We define for brevity  
\begin{equation}
w_\mathrm{X} = \exp{\left(-\tilde{\varepsilon}_\mathrm{X}^2/8\right)},\qquad
w_\mathrm{Y} = \exp{\left(-\tilde{\varepsilon}_\mathrm{Y}^2/8\right)}.
\label{eq:aux}
\end{equation}
Thus, in a strong and in a weak incoherent measurement $w_\mathrm{X},w_\mathrm{Y}\to 0$, while in a weak coherent measurement $w_\mathrm{X},w_\mathrm{Y}\to 1$. 

With these definitions and assumptions,  
the probability \eref{eq:jointprobg} reads
\begin{eqnarray}
\nonumber
\fl
\P\{\Phi,x,y\} = \left|\langle \Phi|\Psi\rangle\right|^2\frac{\varepsilon_\mathrm{X}\varepsilon_\mathrm{Y}}{2\pi}
\biggl\{
|L_\mathrm{w}|^2\,\rme^{-[\varepsilon_\mathrm{X}^2 x_-^{2}+\varepsilon_\mathrm{Y}^2 {y}^{2}]/2}
+\frac{1}{4}\sum_\pm |R_\mathrm{w}\pm \Sigma_\mathrm{w}|^2\,\rme^{-[\varepsilon_\mathrm{X}^2 {x}^{2}+\varepsilon_\mathrm{Y}^2  y_\mp^{2}]/2}
\\
\nonumber
+w_\mathrm{X}w_\mathrm{Y}{{\rme}^{-\varepsilon_\mathrm{X}^2  \left( x-1/2 \right) ^{2}/2}}\sum_\pm  
\mathrm{Re}[L_\mathrm{w}^*(R_\mathrm{w}\pm \Sigma_\mathrm{w})]\,
{{\rme}^{-\varepsilon_\mathrm{Y}^2 \left( y\mp 1/2 \right) ^{2}/2}} 
\\
+\frac{w_\mathrm{Y}^4}{2} \left[|R_\mathrm{w}|^2-|\Sigma_\mathrm{w}|^2\right]{{\rme}^{-\varepsilon_\mathrm{X}^2 {x}^{2}/2-\varepsilon_\mathrm{Y}^2 {y}^{2}/2}}
\biggr\} ,
\label{eq:genjointp}
\end{eqnarray}
We note in the second line of \eref{eq:genjointp} the simultaneous presence of terms corresponding to a shift 
by one half unit, as if half a photon was present in the left and in the right arm.  

The probability of post-selection is 
\begin{equation}
\P\{\Phi\}=\int dx dy \P\{\Phi,x,y\} =\
\left|\langle \Phi|\Psi\rangle\right|^2 \mathcal{N},
\end{equation}
with 
\begin{equation}
\mathcal{N}=
1
-\frac{1}{2}(1-w_\mathrm{Y}^4)\left[|R_\mathrm{w}|^2-|\Sigma_\mathrm{w}|^2\right] 
-2(1-w_\mathrm{X}w_\mathrm{Y})\mathrm{Re}(L_\mathrm{w}^*R_\mathrm{w})	
.
\label{eq:genpostN}
\end{equation}
The conditional probability is therefore 
\begin{eqnarray}
\nonumber
\fl
\P\{x,y|\Phi\} = \frac{\varepsilon_\mathrm{X}\varepsilon_\mathrm{Y}}{2\pi\mathcal{N}}
\biggl\{
|L_\mathrm{w}|^2\,{{\rme}^{-[\varepsilon_\mathrm{X}^2 x_-^{2}+\varepsilon_\mathrm{Y}^2 {y}^{2}]/2}}
+\frac{1}{4}\sum_\pm |R_\mathrm{w}\pm \Sigma_\mathrm{w}|^2\,{{\rme}^{-[\varepsilon_\mathrm{X}^2 {x}^{2}+\varepsilon_\mathrm{Y}^2 y_\mp^{2}]/2}}
\\
\nonumber
+w_\mathrm{X}w_\mathrm{Y}{{\rme}^{-\varepsilon_\mathrm{X}^2  \left( x-1/2 \right) ^{2}/2}}\sum_\pm  
\mathrm{Re}[L_\mathrm{w}^*(R_\mathrm{w}\pm \Sigma_\mathrm{w})]\,
{{\rme}^{-\varepsilon_\mathrm{Y}^2 \left( y\mp 1/2 \right) ^{2}/2}} 
\\
+\frac{w_\mathrm{Y}^4}{2} \left[|R_\mathrm{w}|^2-|\Sigma_\mathrm{w}|^2\right]{{\rme}^{-\varepsilon_\mathrm{X}^2 {x}^{2}/2-\varepsilon_\mathrm{Y}^2 {y}^{2}/2}}
\biggr\} ,
\label{eq:genp}
\end{eqnarray}
and its characteristic function \cite{DiLorenzo2012a} is obtained upon Fourier-transform, 
\begin{eqnarray}
\fl \Z(\chi,\eta)=\Z_0(\chi,\eta)\mathcal{N}^{-1}\biggl\{|L_\mathrm{w}|^2 \rme^{\rmi\chi}+
 \frac{1}{4} \sum_\pm |R_\mathrm{w}\pm\Sigma_\mathrm{w}|^2  \rme^{\pm\rmi\eta}
\nonumber
\\
+w_\mathrm{X}w_\mathrm{Y} \sum_\pm \mathrm{Re}[L_\mathrm{w}^*\left(R_\mathrm{w}\pm\Sigma_\mathrm{w}\right)] \rme^{\rmi(\chi\pm\eta)/2}
 +\frac{w_\mathrm{Y}^4}{2}\left[|R_\mathrm{w}|^2-|\Sigma_\mathrm{w}|^2\right]
\biggr\},
\label{eq:genz}
\end{eqnarray} 
with the initial characteristic function of the detectors 
\begin{equation}
\Z_0(\chi,\eta)= \exp{\left[-\frac{1}{2}\left(\chi^2/\varepsilon_\mathrm{X}^2+\eta^2/\varepsilon_\mathrm{Y}^2\right)\right]} .
\label{eq:charfunc0}
\end{equation}
The appearance of half-counting fields in the third term of the right hand side of \eref{eq:genz} is typical of interference phenomena \cite{DiLorenzo2004,DiLorenzo2006,Forster2005,Urban2008}. 

By differentiating $\Z(\chi,\eta)$ at the origin $\chi=\eta=0$, the moments of the distribution $\P(x,y|\Phi)$ 
are obtained. 
The relevant averages here are 
\mynumparts
\begin{eqnarray}
\langle x\rangle =
\mathcal{N}^{-1}\left[\mathrm{Re}(L_\mathrm{w})-(1-w_\mathrm{X}w_\mathrm{Y})\mathrm{Re}(L_\mathrm{w}^*R_\mathrm{w})\right],\\
\langle y\rangle= 
\mathcal{N}^{-1}\left[\mathrm{Re}(\Sigma_\mathrm{w})-(1-w_\mathrm{X}w_\mathrm{Y}) \mathrm{Re}(L_\mathrm{w}^*\Sigma_\mathrm{w})\right]
,\\
\langle x y\rangle= 
\mathcal{N}^{-1} \frac{w_\mathrm{X}w_\mathrm{Y}}{2}\mathrm{Re}(L_\mathrm{w}^*\Sigma_\mathrm{w}) 
.
\end{eqnarray}
\endmynumparts
In addition to the local averages $\langle x\rangle$ and $\langle y\rangle$, we have written down 
also the cross-average $\langle xy\rangle$. The reason for this is that, while \cite{Aharonov2013} 
considers $\langle x\rangle$ and $\langle y\rangle$ as indicators for a Cheshire cat, in the following 
we shall argue that a better quantifier can be constructed that is proportional to the cross-average 
$\mathcal{C}\propto\langle xy\rangle$. 
\subsection{Both measurements strong or weak incoherent}
\label{subsec:str}
It is assumed that $\tilde{\varepsilon}_\mathrm{X},\tilde{\varepsilon}_\mathrm{Y}\gg1$. 
When $\varepsilon_\mathrm{X},\varepsilon_\mathrm{Y}\gg 1$, the measurement is strong, and the probability of the outputs 
$x,y$ is sharply peaked around the values corresponding to the eigenvalues of the observables $\Pi_\mathrm{L}$ and 
$\sigma_\mathrm{R}$ of the system. Otherwise, if $\varepsilon_\mathrm{X},\varepsilon_\mathrm{Y}\ll 1$, the measurement is weak, in the sense that the peaks are not resolved, but it is incoherent, so that the result can be described in terms of classical ignorance of the initial values $x_0,y_0$ of the pointers.   	
\begin{eqnarray}
\Z(\chi,\eta)=\Z_0(\chi,\eta)\frac{|L_\mathrm{w}|^2 \rme^{\rmi\chi}+
 \frac{1}{4} \sum_\pm |R_\mathrm{w}\pm\Sigma_\mathrm{w}|^2 \rme^{\pm\rmi\eta}}
{|L_\mathrm{w}|^2 +\frac{1}{2} |R_\mathrm{w}|^2+\frac{1}{2}|\Sigma_\mathrm{w}|^2}.
\label{eq:genzstr}
\end{eqnarray} 
The relevant averages are then 
\mynumparts
\begin{eqnarray}
\langle x\rangle =
\frac{|L_\mathrm{w}|^2}{|L_\mathrm{w}|^2 +\frac{1}{2} |R_\mathrm{w}|^2+\frac{1}{2}|\Sigma_\mathrm{w}|^2}
,\\
\langle y\rangle= 
\frac{\mathrm{Re}(R_\mathrm{w}^*\Sigma_\mathrm{w})}
{|L_\mathrm{w}|^2 +\frac{1}{2} |R_\mathrm{w}|^2+\frac{1}{2}|\Sigma_\mathrm{w}|^2},\\
\langle x y\rangle=0 .
\end{eqnarray}
\endmynumparts
\subsection{One measurement strong, the other weak}
First, we note that if the `grin' measurement is strong, $w_\mathrm{Y}\to 0$, all interference effects disappear, and 
we fall back into the case of subsection \ref{subsec:str}. 
We consider instead the limit in which the `cat' measurement is strong, i.e. $w_\mathrm{X}\to 0$, but the `grin' measurement is weak, $w_\mathrm{Y}\to 1$. One would presume that no Cheshire cat is present in this case either. 
In the following, we show 
that it is so, according to our criterion, while the criterion of Aharonov \emph{et al.} instead predicts the 
presence of a Cheshire cat.

From \eref{eq:genpostN} we have the postselection probability 
\begin{eqnarray}
\P\{\Phi\}=\
\left|\langle \Phi|\Psi\rangle\right|^2 \mathcal{N},
\end{eqnarray}
with 
\begin{eqnarray}
\mathcal{N}=
1+2|L_\mathrm{w}|^2-2\mathrm{Re}(L_\mathrm{w})	.
\end{eqnarray}
The characteristic function is 
\begin{eqnarray}
\fl \Z(\chi,\eta)=\Z_0(\chi,\eta)\mathcal{N}^{-1}\biggl\{|L_\mathrm{w}|^2 \rme^{\rmi\chi}+
 \frac{1}{4} \sum_\pm |R_\mathrm{w}\pm\Sigma_\mathrm{w}|^2  \rme^{\pm\rmi\eta}
 +\frac{1}{2}\left[|R_\mathrm{w}|^2-|\Sigma_\mathrm{w}|^2\right]
\biggr\}.
\end{eqnarray} 
The relevant averages are then 
\mynumparts
\begin{eqnarray}
\langle x\rangle =
\mathcal{N}^{-1}|L_\mathrm{w}|^2,\\
\langle y\rangle= 
\mathcal{N}^{-1}\mathrm{Re}(R_\mathrm{w}^* \Sigma_\mathrm{w})
,\\
\langle x y\rangle= 0
.
\end{eqnarray}
\endmynumparts
On one hand, $0\le\langle x \rangle\le1$, with the equality $\langle x \rangle=1$ only if $L_\mathrm{w}=1$, which implies that $\langle y\rangle=0$. 
However, if we allow $\langle x \rangle<1$, we can have a peculiar situation in which a fraction 
$1-\langle x\rangle$ 
of the post-selected photons passes through the right arm (and we have certainty that this is the case, as the measurement revealing the photon is a strong one), yet the amount of polarization $\langle y\rangle$ 
exceeds the maximum amount $1-\langle x\rangle$. If we applied the same reasoning as Aharonov \emph{et al.} \cite{Aharonov2013}, we would erroneously conclude that we have a Cheshire cat. 
However, according to our criterion there is no Cheshire cat, as the absence of interference terms 
between the two arms of the interferometer confirms. 
As an example, let us consider the states
\mynumparts
\begin{eqnarray}
|\Psi\rangle\propto 2|L,+\rangle +2|L,-\rangle
+3|R,+\rangle-2|R,-\rangle,
\\
|\Phi\rangle\propto |L,+\rangle +|L,-\rangle+|R,+\rangle+|R,-\rangle.
\end{eqnarray} 
\endmynumparts
We have that $L_\mathrm{w}=4/5$ and $\Sigma_\mathrm{w}=1$, so that 
$\langle x\rangle = 16/17$ and $\langle y\rangle =5/17$. 
The value $\langle y\rangle$ in excess of the theoretical maximum $1/17$ is due to the 
destructive interference term $(w_\mathrm{Y}^4/2)[|R_\mathrm{w}|^2-|\Sigma_\mathrm{w}|^2]= -12/25$ 
in \eref{eq:genpostN} that 
decreases the probability of post-selection with respect to the one obtained 
for a strong measurement of the `grin', 
from $\P_\mathrm{strong}=29/25|\langle\Phi|\Psi\rangle|^2=29/84$, to 
$\P_\mathrm{weak}=17/25|\langle\Phi|\Psi\rangle|^2=17/84$. This decreased probability, on 
one hand increases the photon count on the left, on the other hand it increases the 
polarization measurement on the right, explaining thus the anomalous value $5/17$. 
We stress how our criterion for the Cheshire cat gives here $\mathcal{C}\propto\langle xy\rangle=0$, 
\subsection{Both measurements weak}
We recall that we are considering the weak coherent measurement, which satisfies 
$\tilde{\varepsilon}_\mathrm{X},\tilde{\varepsilon}_\mathrm{Y}\ll 1$. Since by the uncertainty relation $\varepsilon_\mathrm{X}\le \tilde{\varepsilon}_\mathrm{X}$ and  $\varepsilon_\mathrm{Y}\le \tilde{\varepsilon}_\mathrm{Y}$, this condition implies that the initial uncertainties in the pointer 
variables, $\Delta_x=1/\varepsilon_\mathrm{X},\Delta_y=1/\varepsilon_\mathrm{Y}$ are large. 
Notice that the latter is a necessary but not sufficient condition for the interference effects to manifest, 
as can be seen by inspection of \eref{eq:genp} and \eref{eq:genz}. 
In the limit considered, the characteristic function is 
\begin{eqnarray}
\fl \Z(\chi,\eta)=\Z_0(\chi,\eta)
\biggl\{|L_\mathrm{w}|^2 \rme^{\rmi\chi}+
 \frac{1}{4} \sum_\pm |R_\mathrm{w}\pm\Sigma_\mathrm{w}|^2  \rme^{\pm\rmi\eta}
\nonumber
\\
+ \sum_\pm \mathrm{Re}[L_\mathrm{w}^*\left(R_\mathrm{w}\pm\Sigma_\mathrm{w}\right)] \rme^{\rmi(\chi\pm\eta)/2}
 +\frac{1}{2}\left[|R_\mathrm{w}|^2-|\Sigma_\mathrm{w}|^2\right]
\biggr\}.
\end{eqnarray} 
and the averages are 
\mynumparts
\begin{eqnarray}
\label{eq:wavsa}
\langle x\rangle =
\mathrm{Re}(L_\mathrm{w}),\\
\langle y\rangle= 
\mathrm{Re}(\Sigma_\mathrm{w}),\\
\langle x y\rangle= 
 \frac{1}{2}\mathrm{Re}(L_\mathrm{w}^*\Sigma_\mathrm{w}) 
.
\label{eq:wavsc}
\end{eqnarray}
\label{eq:wavs}
\endmynumparts
We remark the apparent contradiction with section \ref{sec:almorth}: now, if we take the limit 
$\langle \Phi|\Psi\rangle\to 0$, $\langle x\rangle$ and $\langle y\rangle$ diverge. The 
reason of this discrepancy is that one cannot invert the limits $\langle \Phi|\Psi\rangle\to 0$ 
and $w_\mathrm{X},w_\mathrm{Y}\to 1$. Since in any actual experiment the coupling strength is small but finite, so that 
$w_\mathrm{X},w_\mathrm{Y}<1$,  
equations \eref{eq:wavs} are valid as far as $(1-w_\mathrm{Y}^4)\left[|R_\mathrm{w}|^2-|\Sigma_\mathrm{w}|^2\right] 
-4(1-w_\mathrm{X}w_\mathrm{Y})\mathrm{Re}(L_\mathrm{w}^*R_\mathrm{w})\ll 1$. 

\subsection{Discussion}
The preparation and post-selection were chosen by the authors of 
 Ref.~\cite{Aharonov2013} in such a way that, in the weak coupling limit, 
the average outputs take the special values 
$\langle x\rangle =1$ and $\langle y\rangle =1$. 
From this it was inferred that the photon is in the left arm, while its polarization is in the right arm. 
It seems that the following reasoning is implied: 
in a strong measurement, in each individual trial, the output at the right arm is 0, 1, or $-1$, and 
the output at the left arm is either 0 or 1. Thus, if in a strong measurement one would observe an average of 
1, this would imply that the detector always gave the output 1.  
However, this inference does not hold for a weak measurement, as it does not take into account that in each individual repetition,  since the initial spread of the meter is much larger than the readout scale, the observed readout may be  much larger than one (in absolute value). Furthermore, as pointed out in Ref.~\cite{Aharonov1988}, these large values may not cancel out on the average, and they may yield an average output outside the spectrum of the measured observable. 
For instance, one could easily make up a post-selection such 
that $\langle x\rangle =-100$ and 
$\langle y\rangle =100$. Indeed, consider the preparation 
$|\Psi\rangle\propto |L,+\rangle+|L,-\rangle+|R,+\rangle+|R,-\rangle$, 
and the post-selection $|\Phi\rangle\propto |L,+\rangle+|L,-\rangle-0.01|R,+\rangle-2.01|R,-\rangle$. We have then $L_\mathrm{w}=-100$ and $\Sigma_\mathrm{w}=100$. 

While in the example considered in Ref.~\cite{Aharonov1988}, a spin-1/2 system, it may make sense 
to say that the average value of the spin is 100, since from a classical point of view an angular momentum 
has no bounds, the example above illustrates the danger of interpreting the average of a weak measurement 
in a too light-hearted way: it makes no sense, even classically, to say that -100 photons were 
observed by the left detector. 

Even if the average over the post-selected data yields 1, 
this does not warrant the conclusion that all photons passed in the left arm or all were polarized $+$ in the right arm. 
A better way to ascertain whether the polarization is disembodied from the presence of the photon is to 
study the average of the product of the two observables. Indeed, inspection of \eref{eq:genz} 
reveals that only interference terms contribute to the numerator of the cross-average $\langle xy\rangle$, 
while the local averages $\langle x\rangle$ and $\langle y\rangle$, in general, 
contain a contribution from the classical part. We remark that $\langle xy\rangle$ would be zero in a strong or in a weak incoherent measurement, so that its value being nonzero is an indicator of the presence of a quantum Cheshire cat. 
In a realistic experiment, the interaction must be weak enough that 
$\langle xy\rangle$ is larger than the environmental noise in the detectors.  
Furthermore, the common denominator of $\langle x\rangle$, $\langle y\rangle$, and $\langle xy\rangle$,  
$\mathcal{N}\propto \P\{\Phi\}$ contains contributions from both the classical and the interference terms. It can be eliminated by considering the product $\mathcal{C}(\Phi)=\langle xy\rangle \P\{\Phi\}$ as the Cheshire cat parameter. Furthermore, in section \ref{sec:concl} we shall show how 
$\mathcal{C}=2\mathcal{C}(\Phi)$ can be inferred by using all the data, not only the ones where 
the post-selection was successful. 

We remark that the criterion proposed by Aharonov \emph{et al.}, $\langle x\rangle=\langle y\rangle=1$,  would classify as a Cheshire cat situations that, according to our criterion, are not such, and vice versa. 
There may be cases where, e.g., $\langle y\rangle=0$, yet the Cheshire cat is present. 
For instance, in the weak coupling limit, it is sufficient to choose the preparation $|\Psi\rangle$ and the post-selection $|\Phi\rangle$ so that both $L_\mathrm{w}$ and $\Sigma_\mathrm{w}$ are purely imaginary. 
On the other hand, there are cases where, according to Aharonov \emph{et al.} a Cheshire cat is present, 
but according to our criterions it is not. For instance, let us consider a preparation and post-selection in 
pure states such that $L_\mathrm{w}=1+\rmi$ and $\Sigma_\mathrm{w}=1-\rmi$. Then, in the weak 
coupling limit, $\langle x\rangle =1$, $\langle y\rangle =1$, but $\langle x y\rangle =0$. 
These values can be obtained, for instance, with 
$|\Psi\rangle\propto 2|L,+\rangle +2\rmi|L,-\rangle+(1-2\rmi)|R,+\rangle-|R,-\rangle$,
$|\Phi\rangle\propto |L,+\rangle +|L,-\rangle+|R,+\rangle+|R,-\rangle$.

%
\section{Almost orthogonal post-selection and preparation}\label{sec:almorth}
Here, we consider the limit in which $\langle \Phi|\Psi\rangle\ll 1$, so that both $|L_\mathrm{w}|\gg 1$ and 
$|\Sigma_\mathrm{w}|\gg 1$. 
We define the matrix elements 
\mynumparts
\begin{eqnarray}
l_\mathrm{w} = \langle \Phi|\Pi_\mathrm{L}|\Psi\rangle = \langle \Phi|\Psi\rangle L_\mathrm{w},
\\
\sigma_\mathrm{w} = \langle \Phi|\sigma_\mathrm{R}|\Psi\rangle= \langle \Phi|\Psi\rangle \Sigma_\mathrm{w}
.
\end{eqnarray}
\endmynumparts
The post-selection probability, the conditional probability and the characteristic function of the latter are
\begin{eqnarray}
\P\{\Phi\}=\ 2(1-w_\mathrm{X}w_\mathrm{Y})|l_\mathrm{w}|^2
-\frac{1}{2}(1-w_\mathrm{Y}^4)\left[|l_\mathrm{w}|^2-|\sigma_\mathrm{w}|^2\right] 	
,
\label{eq:genppostnopp}
\end{eqnarray}
\begin{eqnarray}
\nonumber
\fl
\P\{x,y|\Phi\} = \frac{\varepsilon_\mathrm{X}\varepsilon_\mathrm{Y}}{2\pi\mathcal{\P}\{\Phi\}}
\biggl\{
|l_\mathrm{w}|^2\,{{\rme}^{-[\varepsilon_\mathrm{X}^2 \left( x-1 \right) ^{2}+\varepsilon_\mathrm{Y}^2 {y}^{2}]/2}}
-\frac{1}{4}\sum_\pm |l_\mathrm{w}\mp \sigma_\mathrm{w}|^2\,{{\rme}^{-[\varepsilon_\mathrm{X}^2 {x}^{2}+\varepsilon_\mathrm{Y}^2  \left( y\mp1\right) ^{2}]/2}}
\\
\nonumber
-w_\mathrm{X}w_\mathrm{Y}{{\rme}^{-\varepsilon_\mathrm{X}^2  \left( x-1/2 \right) ^{2}/2}}\sum_\pm  
\left[ |l_\mathrm{w}| ^2
\mp \mathrm{Re}(l_\mathrm{w}^*\sigma_\mathrm{w})\right]\,
{{\rme}^{-\varepsilon_\mathrm{Y}^2 \left( y\mp 1/2 \right) ^{2}/2}} 
\\
+\frac{w_\mathrm{Y}^4}{2} \left[|l_\mathrm{w}|^2-|\sigma_\mathrm{w}|^2\right]{{\rme}^{-\varepsilon_\mathrm{X}^2 {x}^{2}/2-\varepsilon_\mathrm{Y}^2 {y}^{2}/2}}
\biggr\} ,
\label{eq:genpnopp}
\end{eqnarray}
and 
\begin{eqnarray}
\fl \Z(\chi,\eta)=\frac{\Z_0(\chi,\eta)}{\mathcal{\P}(\Phi)}\biggl\{|l_\mathrm{w}|^2 \rme^{\rmi\chi}-
 \frac{1}{4} \sum_\pm |l_\mathrm{w}\mp\sigma_\mathrm{w}|^2  \rme^{\pm\rmi\eta}
\nonumber
\\
-w_\mathrm{X}w_\mathrm{Y} \sum_\pm \left[|l_\mathrm{w}|^2\mp \mathrm{Re}\left(l_\mathrm{w}^*\sigma_\mathrm{w}\right)\right] \rme^{\rmi(\chi\pm\eta)/2}
 +\frac{w_\mathrm{Y}^4}{2}\left[|l_\mathrm{w}|^2-|\sigma_\mathrm{w}|^2\right]
\biggr\}.
\label{eq:genznopp}
\end{eqnarray} 

The relevant averages are then 
\mynumparts
\begin{eqnarray}
\langle x\rangle =
\P\{\Phi\}^{-1} (1-w_\mathrm{X}w_\mathrm{Y})|l_\mathrm{w}|^2,
\label{eq:noppava}
\\
\langle y\rangle= -\P\{\Phi\}^{-1}(1-w_\mathrm{X}w_\mathrm{Y}) \mathrm{Re}(l_\mathrm{w}^*\sigma_\mathrm{w})
\label{eq:noppavb}
,\\
\langle x y\rangle= \P\{\Phi\}^{-1}\frac{w_\mathrm{X}w_\mathrm{Y}}{2}
\mathrm{Re}(l_\mathrm{w}^*\sigma_\mathrm{w})
.
\label{eq:noppavc}
\end{eqnarray}
\label{eq:noppav}
\endmynumparts
We shall distinguish two subcases: (i) when $\langle \Phi|\Psi\rangle = 0$, the relations 
\eref{eq:genppostnopp}--\eref{eq:noppav} hold exactly for any value of $w_\mathrm{X}$ and $w_\mathrm{Y}$; 
(ii) when $0<|\langle \Phi|\Psi\rangle| = r\ll 1$, the relations 
\eref{eq:genppostnopp}--\eref{eq:noppav} hold as far as 
$r\ll \mathrm{min}(|l_\mathrm{w}|,|\sigma_\mathrm{w}|)$, 
and either $4r^2\ll (3-4w_\mathrm{X}w_\mathrm{Y}+w_\mathrm{Y}^4) |l_\mathrm{w}|^2$ or $4r^2\ll(1-w_\mathrm{Y}^4) |\sigma_\mathrm{w}|^2$.
Thus, in the weak measurement limit, $w_\mathrm{X}\to 1, w_\mathrm{Y}\to 1$, while $\langle x\rangle$ and $\langle y\rangle$ 
stay finite,\footnote{Depending on how this limit is taken, $\langle x\rangle$ and $\langle y\rangle$ take different values, but they never diverge, unless $\sigma_\mathrm{w}=0$. Indeed, let $\phi\in[0,\pi/2]$. Then 
$\langle x\rangle$ and $\langle y\rangle$ can take, in the weak limit, any value of the form 
$\langle x\rangle = (1/2) [(\cos{\phi}+\sin{\phi}) |l_\mathrm{w}|^2]/[\cos{\phi}\,|l_\mathrm{w}|^2 +\sin{\phi}\,|\sigma_\mathrm{w}|^2]$ and 
$\langle y\rangle = (-1/2) [(\cos{\phi}+\sin{\phi}) \mathrm{Re}(l_\mathrm{w}^*\sigma_\mathrm{w})]/[\cos{\phi}\,|l_\mathrm{w}|^2 +\sin{\phi}\,|\sigma_\mathrm{w}|^2]$.}
 the cross-average $\langle xy\rangle$ diverges or becomes of order $1/r^2$, unless 
$\mathrm{Re}(l_\mathrm{w}^*\sigma_\mathrm{w})=0$ . 
%
\section{Mixed state preparation and post-selection}
Our results can be generalized to the case when the system is prepared in a mixed state $\rho_\mathrm{i}$, and it is post-selected in a mixed state $E_\mathrm{f}$. 
While the preparation of a mixture is a commonplace concept, 
how to perform the post-selection in a mixed state is a less well known subject. 
It can be achieved, e.g., by making a positive-operator measurement on the system after 
it has interacted with the detectors \cite{Wiseman2002}, or by making a probabilistic post-selection \cite{DiLorenzo2012a}. 
Strictly speaking, $E_\mathrm{f}$ is a positive operator, the eigenvalues of which do not exceed 1, but which does not necessarily have unit trace. 
Its trace $p_\mathrm{f}=\Tr(E_\mathrm{f})$ is to be interpreted as the prior probability 
of observing an outcome $f$, if nothing further is known about the system, which is thus 
assigned the unpolarized density matrix $\rho_\mathrm{i}\propto\id$. 
While the priors $p_\mathrm{f}$ are positive, they represent 
relative probabilities, in the sense that they are not normalized to one and hence they cannot be 
interpreted as ordinary probabilities. Instead, their sum equals 
the dimension of the system Hilbert space (4, in the present case), and thus it can even diverge. 

From $E_\mathrm{f}$, one can define a density operator $\rho_\mathrm{f}=E_\mathrm{f}/\Tr{E_\mathrm{f}}$.  
The state of the system after 
the post-selection, however, in general does not coincide with $\rho_\mathrm{f}$. 
Instead, $\rho_\mathrm{f}$ represents the optimal state for making retrodictions \cite{Barnett2000,Barnett2001}: if you are told that a system 
was post-selected at time $t_2$ in a state $E_\mathrm{f}$, and that at time $t_1<t_2$ a measurement of an 
observable $\hat{O}$ was made, then, in absence of any further information, the statistics of the measurement are described by $\P(o)=\Tr[\rho_\mathrm{f}(t_1) \Pi_o]$, with $\Pi_o$ the projector on the $o$-th eigenstate of $\hat{O}$ and $\rho_\mathrm{f}(t_1)= U_{t_2,t_1} \rho_\mathrm{f} U_{t_1,t_2}$ the post-selected state evolved back to time $t_1$. 

In general, we need the following weak values:
\mynumparts
\begin{eqnarray}
L_\mathrm{w} = \frac{\Tr[E_\mathrm{f}\Pi_\mathrm{L}\rho_\mathrm{i}]}{\Tr[E_\mathrm{f}\rho_\mathrm{i}]},
\\
\Sigma_\mathrm{w} = 
 \frac{\Tr[E_\mathrm{f}\sigma_\mathrm{R}\rho_\mathrm{i}]}{\Tr[E_\mathrm{f}\rho_\mathrm{i}]},
\\
L_{2\mathrm{w}} = \frac{\Tr[E_\mathrm{f}\Pi_\mathrm{L}\rho_\mathrm{i} \Pi_\mathrm{L}]}{\Tr[E_\mathrm{f}\rho_\mathrm{i}]},
\\
\Sigma_{2\mathrm{w}} = 
 \frac{\Tr[E_\mathrm{f}\sigma_\mathrm{R}\rho_\mathrm{i}\sigma_\mathrm{R}]}{\Tr[E_\mathrm{f}\rho_\mathrm{i}]},
\\
M_\mathrm{w} = 
\frac{\Tr[E_\mathrm{f} \sigma_\mathrm{R}\rho_\mathrm{i}\Pi_\mathrm{L}]}{\Tr[E_\mathrm{f}\rho_\mathrm{i}]}.
\end{eqnarray}
\label{eq:gengenwval}
\endmynumparts
In total, there are 8 real parameters, 3 complex numbers $L_\mathrm{w},\Sigma_\mathrm{w},M_\mathrm{w}$ and 
2 positive real numbers $L_{2\mathrm{w}},\Sigma_{2\mathrm{w}}$. 
For pure preparation and post-selection, $L_{2\mathrm{w}}=|L_\mathrm{w}|^2$, $\Sigma_{2\mathrm{w}}=|\Sigma_\mathrm{w}|^2$, 
and $M_\mathrm{w}=L_\mathrm{w}^*\Sigma_\mathrm{w}$.   
In order to keep equations relatively simple, we shall define also some redundant 
weak values
\mynumparts
\begin{eqnarray}
R_\mathrm{w} =  \frac{\Tr[E_\mathrm{f}\Pi_\mathrm{R}\rho_\mathrm{i}]}{\Tr[E_\mathrm{f}\rho_\mathrm{i}]}=1-L_\mathrm{w},
\\
R_{2\mathrm{w}} = \frac{\Tr[E_\mathrm{f}\Pi_\mathrm{R}\rho_\mathrm{i} \Pi_\mathrm{R}]}{\Tr[E_\mathrm{f}\rho_\mathrm{i}]}=1-2\mathrm{Re}(L_\mathrm{w})+L_{2\mathrm{w}},
\\
Q_\mathrm{w} = 
\frac{\Tr[E_\mathrm{f} \Pi_\mathrm{L}\rho_\mathrm{i}\Pi_\mathrm{R}]}{\Tr[E_\mathrm{f}\rho_\mathrm{i}]}
= L_\mathrm{w}-L_{2\mathrm{w}},
\\
N_\mathrm{w} = 
\frac{\Tr[E_\mathrm{f} \sigma_\mathrm{R}\rho_\mathrm{i}\Pi_\mathrm{R}]}{\Tr[E_\mathrm{f}\rho_\mathrm{i}]}=\Sigma_\mathrm{w}-M_\mathrm{w},
\end{eqnarray}
\label{eq:gengenaddwval}
\endmynumparts
The post-selection probability is  
\begin{eqnarray}
\P\{E_\mathrm{f}\}=\
\Tr[E_\mathrm{f}\rho_\mathrm{i}] \mathcal{N}
\label{eq:gengenpost}
\end{eqnarray}
with 
\begin{eqnarray}
\mathcal{N}=
1
-\frac{1}{2}(1-w_\mathrm{Y}^4)\left[R_{2\mathrm{w}}-\Sigma_{2\mathrm{w}}\right] 
-2(1-w_\mathrm{X}w_\mathrm{Y})\mathrm{Re}(Q_\mathrm{w})
.
\label{eq:gengenN}
\end{eqnarray}
The conditional probability is 
\begin{eqnarray}
\fl
\P\{x,y|E_\mathrm{f}\} \!=\! \frac{\varepsilon_\mathrm{X}\varepsilon_\mathrm{Y}}{2\pi\mathcal{N}}
\biggl\{\!L_{2\mathrm{w}}\,{{\rme}^{-[\varepsilon_\mathrm{X}^2 x_-^{2}+\varepsilon_\mathrm{Y}^2 {y}^{2}]/2}}
+\sum_\pm \frac{R_{2\mathrm{w}}+ \Sigma_{2\mathrm{w}}\pm 2\mathrm{Re}(N_\mathrm{w})}{4}\,{{\rme}^{-[\varepsilon_\mathrm{X}^2 {x}^{2}+\varepsilon_\mathrm{Y}^2  y_\mp^{2}]/2}}
\nonumber
\\
+w_\mathrm{X}w_\mathrm{Y}{{\rme}^{-\varepsilon_\mathrm{X}^2  \left( x-1/2 \right) ^{2}/2}}
\sum_\pm  \mathrm{Re}\left(Q_\mathrm{w}\pm M_\mathrm{w}\right)\,
{{\rme}^{-\varepsilon_\mathrm{Y}^2 \left( y\mp 1/2 \right) ^{2}/2}} 
\nonumber
\\
+\frac{w_\mathrm{Y}^4}{2} \left[R_{2\mathrm{w}}-\Sigma_{2\mathrm{w}}\right]{{\rme}^{-\varepsilon_\mathrm{X}^2 {x}^{2}/2-\varepsilon_\mathrm{Y}^2 {y}^{2}/2}}
\biggr\} ,
\label{eq:gengenp}
\end{eqnarray}
and its characteristic function is 
\begin{eqnarray}
\fl \Z(\chi,\eta)=\Z_0(\chi,\eta)\mathcal{N}^{-1}\biggl\{L_\mathrm{2w} \rme^{\rmi\chi}+
 \frac{1}{4} \sum_\pm \left[R_{2\mathrm{w}}+ \Sigma_{2\mathrm{w}}\pm 2\mathrm{Re}(N_\mathrm{w})\right] \rme^{\pm\rmi\eta}
\nonumber
\\
+w_\mathrm{X}w_\mathrm{Y} \sum_\pm \mathrm{Re}\left(Q_\mathrm{w} \pm M_\mathrm{w}\right) \rme^{\rmi(\chi\pm\eta)/2}
 +\frac{w_\mathrm{Y}^4}{2}\left[R_{2\mathrm{w}}-\Sigma_{2\mathrm{w}}\right]
\biggr\}.
\label{eq:gengenz}
\end{eqnarray} 
The relevant averages are then 
\mynumparts
\begin{eqnarray}
\langle x\rangle =
\mathcal{N}^{-1}\left[w_\mathrm{X}w_\mathrm{Y} \mathrm{Re}(L_\mathrm{w})+(1-w_\mathrm{X}w_\mathrm{Y})L_\mathrm{2w}\right],\\
\langle y\rangle= 
\mathcal{N}^{-1}\left[\mathrm{Re}(\Sigma_\mathrm{w})-(1-w_\mathrm{X}w_\mathrm{Y}) \mathrm{Re}(M_\mathrm{w})\right]
,\\
\langle x y\rangle= 
\mathcal{N}^{-1} \frac{w_\mathrm{X}w_\mathrm{Y}}{2}\mathrm{Re}(M_\mathrm{w}) 
.
\end{eqnarray}
\label{eq:gengenavs}
\endmynumparts

In particular, when no post-selection is made, i.e. $E_\mathrm{f}\propto \id$, or 
when the preparation is in the completely unpolarized state $\rho_\mathrm{i}\propto\id$, 
then $M_\mathrm{w}=Q_\mathrm{w}=0$, so that the interference terms disappear and in particular
$\langle xy\rangle=0$. 
Furthermore, the Cheshire cat indicator vanishes also if either the preparation or the post-selection is in a mixture of states localized in the left and in the right arm, i.e. if 
\begin{equation}
\rho_\mathrm{i}\ \mathrm{ or }\  E_\mathrm{f}= 
\left(\begin{array}{cc}
P_L&0_2\\
0_2&P_R
\end{array}\right),
\end{equation}
with $P_L$ and $P_R$ being $2\times2$ non-negative matrices, 
and $0_2$ being $2\times2$ null matrices. 
This behavior confirms the validity of our criterion for characterizing a Cheshire cat state.  
%
\section{Conclusions}
\label{sec:concl}
It is clear from the general results of the preceding sections that the term 
\begin{equation}
\mathcal{C}(E_\mathrm{f})	=\langle xy\rangle \P\{E_\mathrm{f}\}
\label{eq:expc}
\end{equation}
provides the signature 
for the quantum Cheshire cat, i.e. for the interference between the presence of the photon 
and its polarization in the two arms of the interferometer. 
\Eref{eq:expc} is the operational definition of the Cheshire cat parameter, as it can be obtained 
directly from experimental quantities. 
The theoretical value, on the other hand,  is 
\begin{equation}
\mathcal{C}(E_\mathrm{f})=\frac{w_\mathrm{X}w_\mathrm{Y}}{2}\mathrm{Re}\left[\Tr{(E_\mathrm{f}\sigma_\mathrm{R}\rho_\mathrm{i}\Pi_\mathrm{L})}\right].
\label{eq:thc}
\end{equation}
While post-selection is a necessary ingredient for the observation of a Cheshire cat, 
as interference terms disappear in its absence, one needs not retain only the data corresponding 
to a successful post-selection. Indeed, when the post-selection in the state $E_\mathrm{f}$ fails, 
the system is post-selected in the complementary state $E'_\mathrm{f}=\id-E_\mathrm{f}$. 
Therefore, for an unsuccessful post-selection
\begin{equation}
\mathcal{C}(E'_\mathrm{f})=\frac{w_\mathrm{X}w_\mathrm{Y}}{2}\mathrm{Re}\left\{\Tr{[(\id-E_\mathrm{f})\sigma_\mathrm{R}\rho_\mathrm{i}\Pi_\mathrm{L}]}\right\} = -\mathcal{C}(E_\mathrm{f}),
\label{eq:thc2}
\end{equation}
where we exploited the cyclic property of the trace and used $\Pi_\mathrm{L}\sigma_\mathrm{R}=0$. 
The following operational prescription follows: 
In the $j$-th trial, if the post-selection is successful, consider the product $c_j=x_j y_j$, 
otherwise, consider $c_j=-x_j y_j$; sum the $c_j$ and divide by the number of trials; 
the value $\mathcal{C}=2\mathcal{C}(E_\mathrm{f})$ is thus obtained, allowing to establish whether a Cheshire cat is observed or not. 

The maximum value of the Cheshire cat parameter is $\mathcal{C}_\mathrm{max}=w_\mathrm{X}w_\mathrm{Y}/4$, and it is attained for 
the preparation and postselection
\begin{eqnarray}
|\Psi\rangle = \frac{a}{\sqrt{2}}|L,+\rangle+\frac{b}{\sqrt{2}}|L,-\rangle+e^{i\phi}\frac{1}{2}|R,+\rangle
+\frac{1}{2}|R,-\rangle,
\\
|\Phi\rangle = \frac{a}{\sqrt{2}}|L,+\rangle+\frac{b}{\sqrt{2}}|L,-\rangle+e^{i\phi}\frac{1}{2}|R,+\rangle
-\frac{1}{2}|R,-\rangle,
\end{eqnarray}
with $|a|^2+|b|^2=1$ and $\phi\in[0,2\pi]$ an arbitrary phase. 
Furthermore, the criterion $\mathcal{C}\neq 0$ does not require a very weak coupling, 
as an intermediate one is sufficient, provided that the signal is above the external noise.
We stress, indeed, that we have so far assumed that 
the readout of the detectors is projective and error-less, the only uncertainty $\Delta_x,\Delta_y$ coming 
from the initial preparation of the meters. When external noise is accounted for, let 
us call its square variance $\nu_x,\nu_y$, the criterion to observe unambiguously a Cheshire cat is that 
$\nu_x\nu_y\ll \mathcal{C}$. A necessary condition is that $\nu_x\ll w_\mathrm{X}/2$ and $\nu_y\ll w_\mathrm{Y}/2$.

Finally, it may happen that $\mathcal{C}= 0$ because $\mathrm{Re}(L_\mathrm{w}^*\Sigma_\mathrm{w})=0$ (or, more generally, because $\mathrm{Re}(M_\mathrm{w})=0$), 
but interference is still present. In this case, inspection of \eref{eq:gengenz} shows that 
the simplest quantity to consider is 
\begin{equation}
\langle xy^2\rangle = \langle x\rangle \Delta_y^2 + 
 \frac{w_\mathrm{X}w_\mathrm{Y}}{4\mathcal{N}}\mathrm{Re}(Q_\mathrm{w}) 
.
\end{equation}
If this cross-average differs from $\langle x\rangle \Delta_y^2$, then interference effects are 
detectable. However, we would not call such a term a Cheshire cat, for the simple reason that the 
polarization of the photon does not enter in it. Rather, this term reveals how the cat is 
simultaneously present in both arms.


\ack
This work was performed as part of the Brazilian Instituto Nacional de Ci\^{e}ncia e
Tecnologia para a Informa\c{c}\~{a}o Qu\^{a}ntica (INCT--IQ), it
was supported by Funda\c{c}\~{a}o de Amparo \`{a} Pesquisa do 
Estado de Minas Gerais through Process No. APQ-02804-10 and 
by the Conselho Nacional de Desenvolvimento Cient\'{\i}fico e Tecnol\'{o}gico (CNPq) 
through Process no. 245952/2012-8.

\section*{References}
\providecommand{\newblock}{}

\end{document}